# Cost/Benefit Analysis Model for Implementing Virtual Reality in Construction Companies

Payam Mohammadi[1]
Claudia Garrido Martins, Ph.D., PMP[2]

[1]Ph.D. Student, Dept. of Engineering Technology and Construction Management, University of North Carolina at Charlotte (Corresponding Author) Email: pmohamm1@charlotte.edu
[2]Assistant Professor, Dept. of Engineering Technology and Construction Management, University of North Carolina at Charlotte. Email: cgarrido@charlotte.edu

**ABSTRACT**

Immersive technologies (ImT), like Virtual Reality (VR), have several potential applications in the construction industry. However, the absence of a cost-benefit analysis discourages construction decision-makers from implementing these technologies. In this study, we proposed a primary model for conducting a cost-benefit analysis for implementing virtual reality in construction companies. The cost and benefit factors were identified through a literature review and considered input variables for the model, and then using synthetic data, a Monte Carlo simulation was performed to generate a distribution of outcome. Given the uncertainty in input parameters, this distribution reflected the potential range of total net benefit. Considering synthetic data and input factors obtained only through literature and assumptions, VR implementation could be a promising decision based on the results. This study's results would benefit decision-makers in construction companies about the costs and benefits of implementing VR and other researchers interested in this field.

**INTRODUCTION**

The importance of VR technology in the Architecture, Engineering, and Construction (AEC) sectors should not be underestimated. The potential applications of VR in the construction sector encompass stakeholder engagement, design assistance and evaluation, progress tracking, employee training, and other relevant domains (Albahbah et al. 2021, Baeza 2018, Davila Delgado et al. 2020a, b, Nassereddine et al. 2022, Piroozfar et al. 2017). Despite the significant promise offered by VR, its use within the (AEC) sector continues to be less than optimum (Davila Delgado et al. 2020a). (Dodge Data & Analytics 2022) ran an online survey to collect data from contractors and engineers actively engaged in civil projects. Based on their results, VR and Augmented Reality (AR) technologies implementation within the civil contracting industry are said to be below 10%. Moreover, there is a lack of empirical data indicating small-scale contractors' adoption of these technologies.

The broad use of these technologies is sometimes hindered by the perceived obstacles of high implementation costs and a lack of apparent benefits (Davila Delgado et al. 2020a, b, Nassereddine et al. 2022). In fact, the individuals responsible for making decisions seek to ascertain the potential return on investment associated with implementing immersive technologies. To address this inquiry, (Oesterreich and Teuteberg 2017) developed a comprehensive assessment methodology for evaluating the costs and benefits associated with investments in information systems (IS). The framework is grounded in an application scenario of AR inside the construction sector. The assessment of the subject was conducted through a simulation employing the Visualization of Financial Implications (VoFI) appraisal methodology. (Pedram et al. 2021) conducted a study that examined the utilization of VR as a means of safety training for emergency rescue workers in the mining industry. The technique utilized is based on the socio-technical system theory. This methodology allows participants to provide immediate feedback, which is subsequently used to evaluate the social subsystem costs and benefits of VR training technology. In another study, (Chen et al. 2019) investigated the costs and benefits associated with visualization in virtual environments. The study specifically examined these aspects from three distinct perspectives: information theory, cognitive sciences, and practical applications.

It is important to acknowledge that despite thoroughly examining the available literature, no prior contributions were found that specifically addressed the financial ramifications associated with the use of VR technology within the construction sector. Indeed, a significant portion of the existing research has acknowledged the considerable costs associated with VR as a primary impediment to its extensive adoption within the construction industry. However, the specific components contributing to this elevated cost have not been disclosed. It is crucial for decision-makers to possess a comprehensive understanding of the costs and benefits of implementation in a quantifiable manner. Given the significance of this knowledge gap, there arises an evident urgency for the conceptualization and development of a robust cost-benefit analysis model or framework that pertains specifically to the integration of VR. In this study, our aim is to bridge this gap by proposing a preliminary model rooted in the principle of Monte Carlo simulation. This methodology promises to provide a comprehensive perspective on the range of potential outcome (total net benefit), accounting for the inherent uncertainties within the construction industry and the input variables for the model (Cost and Benefit factors). By deploying this model, we endeavor to offer construction stakeholders, researchers, and practitioners an invaluable tool that not only demystifies the financial dimensions but also aids in informed decision-making concerning the integration of VR within the construction sector.

**METHODOLOGY**

Figure 1 depicts a flow diagram of the methodology applied to this study, including identifying and categorizing cost-benefit factors, monetizing these factors to provide the input variables for the model, and finally, cost-benefit analysis and developing the model. A Monte Carlo simulation was performed using the @Risk add-in for Microsoft Excel developed by Palisade (Lumivero) Corporation for model development.

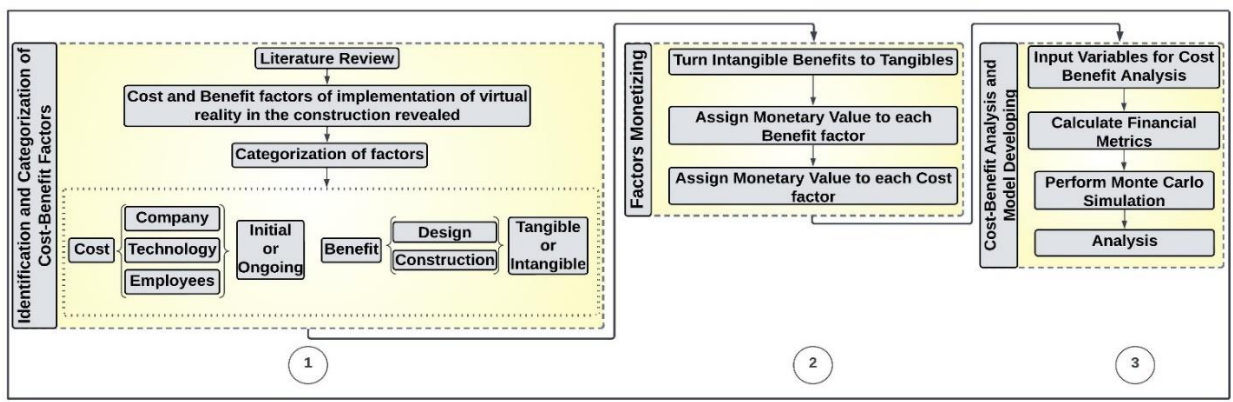

**Figure 1. Methodology Phases**

**Phase 1.** *Identification and Categorization of Cost-Benefit Factors.* Several databases, such as Google Scholar, Science Direct, and Web of Science, were used to collect relevant journal articles, conference papers, and construction industry reports to find potential cost or benefit factors. The search was limited between 2003 and 2023 to cover 20 years. In total, five keywords related to virtual reality were used to collect documents, including 1) Virtual Reality, 2) VR, 3) Immersive Technologies, 4) Cost, and 5) Benefit. Records that matched at least one of the Virtual Reality, VR, or Immersive Technologies terms and one of the Cost or Benefit terms in their title, keywords, or abstract appeared in the search results. The 238 records that met all search criteria were chosen for screening. During the review, the goal was to look for documents that gave information about the construction industry or construction companies. In the end, after screening all the documents, 100 records were selected as references which mentioned the costs or benefits of implementation of virtual reality in the construction sector. In total, 22 benefits were identified, and 15 were used in the model because the remaining seven were considered factors unrelated to the construction companies' daily operation, like "*providing a way to operate the facility remotely in an immersive environment.*"

In addition, the existing body of knowledge referencing the cost taxonomies outlined in the Information Systems literature was utilized. We have opted to use the cost taxonomies proposed by (Irani et al. 2006). To accomplish this task, the taxonomies were integrated into a revised cost taxonomy, as presented in Table 1. In addition, due to the inherent similarities between AR and VR, some other cost factors were also derived from literature related to the implementation of AR. In total, 17 cost factors were identified, and 14 were used in the model for the same reason of eliminating some benefits.

**Table 1. Cost-Benefit Factors**

| Number | Cost Factors | Reference |
|---|---|---|
| 1 | Costs of hardware (HMDs, Smart Glasses, sensors) | (Adebowale and Agumba 2022, Van Krevelen and Poelman 2010) |
| 2 | Costs of software (Apps, contents, licenses, etc.) | (Adebowale and Agumba 2022, Badamasi et al. 2022, Irani et al. 2006) |
| 3 | Costs of consulting | (Irani et al. 2006, Van Krevelen and Poelman 2010) |
| 4 | Costs of providing Infrastructure (design of appropriate office, internet connection) | (Irani et al. 2006) |
| 5 | Business process re-engineering (BPR) expenses | (Irani et al. 2006) |
| 6 | Training expenses for personnel who will use the application | (Davila Delgado et al. 2020a, Irani et al. 2006) |
| 7 | Costs of hardware maintenance | (Irani et al. 2006) |
| 8 | Costs of software modifications | (Irani et al. 2006) |
| 9 | Cost of hardware modifications | (Irani et al. 2006) |
| 10 | Rental expenses | (Irani et al. 2006) |
| 11 | Support cost (the expense of internal/external support required in the event of faults or damage to either the software or the hardware) | (Irani et al. 2006) |

| 12 | Change management costs | (Irani et al. 2006) |
| 13 | Overheads (Insurance, Electricity Consumption, Internet Connection) | (Irani et al. 2006) |
| 14 | Cost of management and staff dealing with procurement | (Irani et al. 2006) |
| **Number** | **Benefit Factors** | |
| 1 | Increase in potential client engagement | (Grudzewski et al. 2018, Juan et al. 2018) |
| 2 | Support designers in identifying the consequences of their design decisions | (Natephra et al. 2017) |
| 3 | Facilitate the communication of design intent | (Berg and Vance 2017) |
| 4 | Visualizes CAD data and allows for an intuitive interaction | (Setareh et al. 2005) |
| 5 | Provides a systematic way for architectural design firms to get accurate user feedback | (Ergan et al. 2018) |
| 6 | Support constructability analysis meetings | (Boton 2018) |
| 7 | Simulate complex construction operations | (Turner et al. 2016) |
| 8 | Carry out progress monitoring remotely | (Robbins et al. 2009) |
| 9 | Provide safer working environments by contributing to hazard identification | (Moore and Gheisari 2019) |
| 10 | Support safety education | (Moore and Gheisari 2019) |
| 11 | Evaluates the lift operation in real-time in terms of its safety and practicability | (Pooladvand et al. 2021) |
| 12 | Improve ironworkers' productivity performance | (Teizer et al. 2013) |
| 13 | Effective in creating teleoperation technologies and quantitatively evaluating them | (Kamezaki et al. 2013) |
| 14 | Enabled engineers to complete their tasks after a brief training session | (Su et al. 2013) |
| 15 | Owners have a better understanding of their projects and thus make timely decisions | (Koutsabasis et al. 2012) |

After identifying factors in the previous step, they were categorized into five categories based on their inherent nature. For instance, for cost factors, they were classified as **Company-related** costs (e.g., cost of providing infrastructure and overheads), **Technology** related costs (e.g., cost of purchasing hardware and software), and **Employee** related costs (e.g., training expenses and support costs). For benefit factors, they were categorized as benefits associated with the **Design** phase (e.g., visualizing CAD data or supporting designers) and benefits related to the **Construction** phase (e.g., simulating complex construction operations or improving ironworkers' performance). Also, cost factors were grouped into **Initial** (necessary for launching the new technology implementation and only once occurs at the beginning) and **Ongoing** (that will happen each year to maintain the technology implementation smoothly and continuously). In addition, benefit factors were grouped into **Tangible** (benefits that could easily be monetized) and **Intangible** (benefits that are hard to monetize).

**Phase 2.** *Factors Monetizing.* The challenge of quantifying benefits is widely regarded as the primary impediment to conducting a cost-benefit analysis, owing to the diverse array of intangible benefits. Measurement issues arise specifically from quantifying intangible benefits such as "improved client engagement" or "acquiring user feedback," as it is challenging to attribute values to these variables. Therefore, as a solution, these benefits were decomposed into smaller components to quantify them. Eight of the fifteen benefits used in the model were intangibles, which were required to turn into measurable terms. For instance, "*It can support designers in identifying the consequences of their design decisions and having a better understanding of the final results*" was re-expressed as "*Reduced design errors and rework.*"

In this preliminary model, the synthetic data were used to provide monetary values for input variables (cost and benefit factors). For each benefit factor, a range of possible minimum and maximum values were assumed for a single project per year. For instance, for the benefit "*increase in potential client engagement*," we assumed that by leveraging VR to provide a realistic representation of a built asset, construction companies could potentially attract more clients and engage them in the decision-making process, which can result in more client satisfaction and higher revenue. The impact on revenue would depend on factors such as the project size and the number of potential clients engaged. Then, assuming a Gamma distribution, the final monetary value was calculated for a minimum of $10,000 and a maximum of $100,000 (=RiskGamma (10000,100000)), and then the final value was inserted in the model ($54,842). This value was considered for the first year after the implementation of the VR, and a 5% increase was assumed for the following years until the end of year 5. For cost factors, for a few of them, the average price in the current market was considered. For instance, for the cost factor (*cost of hardware*), each required hardware's average price in the market in February of 2023 was obtained through an online search and added together to reach the final value as follows: headset and motion controllers = $750, high-performance computers = $1364, and tracking devices = $130 (Total of $2,244). For other cost factors, assuming a normal distribution, their final value was calculated exactly like benefit factors. Table 2 shows the cost and benefit monetary values in 5 years.

Table 2. Cost and Benefit Monetary Values.

| Initial Cost Factors | Time Period | | | | | |
|---|---|---|---|---|---|---|
| | 0 | 1 | 2 | 3 | 4 | 5 |
| 1 | $2,244 | | | | | |
| 2 | $5,250 | | | | | |
| 3 | $5,500 | | | | | |
| 4 | $31,500 | | | | | |
| 5 | $82,500 | | | | | |
| 6 | $5,250 | | | | | |
| Total Initial Costs | $132,244 | | | | | |
| Ongoing Cost Factors | | | | | | |
| 2 | | $2,438 | $2,560 | $2,688 | $2,822 | $2,963 |
| 3 | | $5,500 | $5,775 | $6,064 | $6,376 | $6,685 |
| 7 | | $470 | $494 | $518 | $544 | $571 |
| 8 | | $2,650 | $2,783 | $2,922 | $3,068 | $3,221 |
| 9 | | $1,750 | $1,838 | $1,929 | $2,026 | $2,127 |
| 10 | | $630 | $662 | $695 | $729 | $766 |
| 11 | | $200 | $210 | $221 | $232 | $243 |
| 12 | | $15,000 | $15,750 | $16,538 | $17,364 | $18,233 |
| 13 | | $12,650 | $13,283 | $13,947 | $14,644 | $15,376 |
| 14 | | $125,000 | $131,250 | $137,813 | $144,703 | $151,938 |
| **Total Yearly Ongoing Costs** | $0 | $166,288 | $174,605 | $183,335 | $192,508 | $202,123 |
| **Total Ongoing Costs** | $918,859 | | | | | |
| **Total Costs** | **$1,051,103** | | | | | |
| **Benefit Factors** | **Time Period** | | | | | |
| | 0 | 1 | 2 | 3 | 4 | 5 |
| 1 | | $54,842 | $57,584 | $60,463 | $63,486 | $66,661 |
| 2 | | $27,507 | $28,882 | $30,326 | $31,843 | $33,435 |
| 3 | | $15,972 | $16,771 | $17,609 | $18,490 | $19,414 |
| 4 | | $12,481 | $13,105 | $13,760 | $14,448 | $15,171 |
| 5 | | $12,481 | $13,105 | $13,760 | $14,448 | $15,171 |
| 6 | | $12,481 | $13,105 | $13,760 | $14,448 | $15,171 |
| 7 | | $27,507 | $28,882 | $30,326 | $31,843 | $33,435 |
| 8 | | $17,476 | $18,350 | $19,267 | $20,231 | $21,242 |
| 9 | | $27,507 | $28,882 | $30,326 | $31,843 | $33,435 |
| 10 | | $27,507 | $28,882 | $30,326 | $31,843 | $33,435 |
| 11 | | $27,507 | $28,882 | $30,326 | $31,843 | $33,435 |
| 12 | | $17,476 | $18,350 | $19,267 | $20,231 | $21,242 |
| 13 | | $27,507 | $28,882 | $30,326 | $31,843 | $33,435 |
| 14 | | $27,507 | $28,882 | $30,326 | $31,843 | $33,435 |
| 15 | | $50,987 | $53,536 | $56,213 | $59,024 | $61,975 |
| **Total Yearly Benefit** | $0 | $386,745 | $406,080 | $426,381 | $447,707 | $470,092 |
| **Total Benefits** | **$2,137,005** | | | | | |

**Phase 3.** *Cost-Benefit Analysis and Model Developing.*
In the next step, financial metrics such as Cash Flows and Net Present Value were calculated, and then a Monte Carlo simulation was performed using the @Risk add-in. The input variables of this simulation were Total Benefits and Total Costs, and our output variable of interest was Total Net Benefits. An automatic number of runs was considered to run the simulation by setting the convergence tolerance to 1% and a confidence level of 95%.

**RESULTS & DISCUSSION**

Through quantitative analysis, risk assessment, consistency enforcement, and effective communication, integrating models, particularly cutting-edge methods like Monte Carlo simulation for cost-benefit analysis, significantly improves the decision-making process. By transforming qualitative factors into quantifiable data and enabling the objective comparison of alternatives, these models provide a structured framework for assessing complex scenarios. Furthermore, models like the one proposed in this study promote reliability and consistency by reducing bias, and their transparency makes it easier to comprehend how results came to be. In this simulation, as shown in Table 2, based on synthetic data, by investing the total amount of **$1,051,103,** including an initial investment of **$132,244** and the total ongoing costs of **$918,859** in 5 years, the company could benefit from a total amount of **$2,137,005**. On the other hand, table 3 shows the important financial metrics for decision-making. For instance, the investment's Net Present Value (NPV) equals **$917,560**.

Table 3. Financial metrics

| Financial Metrics | Time Period | | | | | |
|---|---|---|---|---|---|---|
| | 0 | 1 | 2 | 3 | 4 | 5 |
| **Cash Inflow** | 0 | $386,745 | $406,080 | $426,381 | $447,707 | $470,092 |
| **Cash Outflow** | $132,244 | $166,288 | $174,605 | $183,335 | $192,508 | $202,123 |
| **Net Cash Flow** | -$132,244 | $220,457 | $231,475 | $243,046 | $255,199 | $267,969 |
| **Cumulative Net Cash Flow** | -$132,244 | $88,213 | $319,688 | $562,734 | $817,933 | $1,085,902 |
| **Net Present Value** | $917,560 | | | | | |
| **Total Net Benefit** | $1,085,902 | | | | | |

Also, Figure 2 depicts the result of the Monte Carlo simulation. The simulation outcome gave us the Mean of **$1,086,770.12** with a standard deviation of **$312,136.23**. Also, the mode is **$1,210,178.28**.

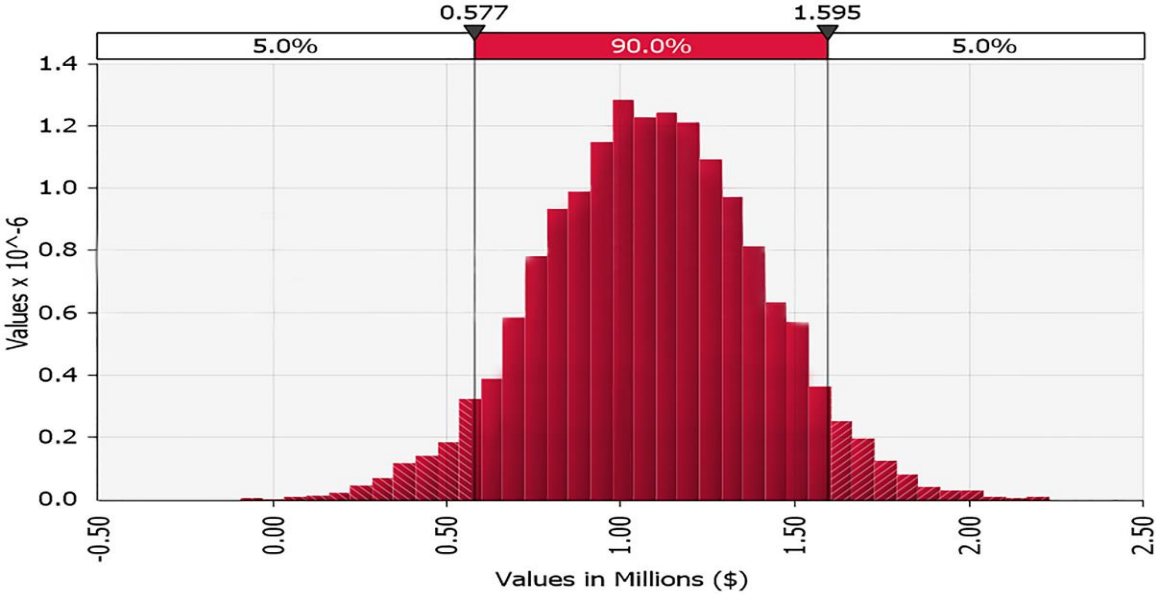

Figure 2. Monte Carlo Simulation

## CONCLUSION

This research sought to establish a preliminary cost-benefit analysis model for implementing virtual reality in construction companies. The synthetic data were used to provide values for input variables. The results indicate that in 5 years, the implementation of VR could be a reasonable investment for construction companies. Also, the following factors could be mentioned as some important factors to be considered by the companies in their decision-making process besides the cost and benefit factors: the type of project, the phase of the desired project (design or construction), the type of technology they want to implement (whether it is head-mounted VR or desktop VR), and the type of VR environment development (whether it is internal or external). Construction companies could use the proposed model in this study to insert any other input factors to see their desired outcomes under different conditions.

A future step would be the investigation of construction companies to collect their experience and data regarding the implementation of VR and its costs and benefits for future model development. Also, performing a sensitivity analysis would be another necessary future step. The analysis considers the range of possible values for each input and calculates the probability of outcome values.